\documentclass[preprint]{aastex}

\def\gta{\;\lower 0.5ex\hbox{$\buildrel > \over \sim\ $}}
\def\lta{\;\lower 0.5ex\hbox{$\buildrel < \over \sim\ $}}
\def\arcsec{$^{\prime\prime}$}
\def\arcmin{$^{\prime}$}
\def\deg{$^{\circ}$}
\def\etal{{\it\ et al.}}
\def\ie{{\it i.e.}}
\def\eg{{\it e.g.}}
\def\qv{{\it q.v.}}

\def\kms{\ km s$^{-1}$}
\def\HII{H${\scriptstyle\rm II}$}

\begin{document}

\title{The Large-scale Bipolar Wind in the Galactic Center}

\author{Joss Bland-Hawthorn}
\affil{Anglo-Australian Observatory, Epping, NSW, Australia}

\author{Martin Cohen}
\affil{Radio Astronomy Laboratory, 601 Campbell Hall, Univ. of California, Berkeley, CA}

\begin{abstract}
During a 9-month campaign (1996--1997), the Midcourse Space Experiment
(MSX) satellite mapped the Galactic Plane at mid-infrared wavelengths
(4.3--21.3$\mu$m).  The much greater spatial resolution and sensitivity
of the MSX 8.3$\mu$m band compared to IRAS at 12$\mu$m has revealed
diffuse emission not seen before. Here we report evidence for a
spectacular limb-brightened, bipolar structure at the Galactic Center
extending more than a degree (170~pc at 8.0~kpc) on either side of the
plane. The 8.3$\mu$m filamentary emission appears confined to a
dome-shaped shell which extends perpendicular to the plane, and shows a
tight correlation with the 3, 6 and 11~cm continuum emission over the
same scales.  The most likely scenario is that the extraplanar
8.3$\mu$m emission arises from dust entrained in a large-scale
bipolar wind powered by a central starburst.  The inferred energy
injection at the source is $\sim$10$^{54}/\kappa$ erg for which
$\kappa$ is the covering fraction of the dusty shell ($\kappa \lta
0.1$).  In accord with the stellar record in the Galactic Center, we
infer that a powerful nuclear starburst has taken place within the last
several million years.  The inferred energies require $\gta
10^4/\varepsilon$ supernovae, where $\varepsilon$ is the thermalization
efficiency of the supernova ejecta.

There is observational evidence for a galactic wind on much larger
scales, presumably from the same central source which produced the
bipolar shell seen by MSX.  Sofue has long argued that the North Polar
Spur -- a thermal x-ray/radio loop which extends from the Galactic
Plane to $b = +80$\deg -- was powered by a nuclear explosion
($1-30\times 10^{55}$ erg) roughly 15~Myr ago, although his published
models did not explain the projected loop structure.  We demonstrate
that an open-ended bipolar model when viewed in projection {\it in the
near field} provides the most natural explanation for the observed loop
structure. The ROSAT 1.5~keV diffuse x-ray map over the inner
45\deg\ provides compelling evidence for this interpretation. Since the 
faint bipolar emission would be very difficult to detect beyond the
Galaxy, {\it the phenomenon of large-scale galactic winds may be far more 
common than has been observed to date.} The derived energetics for the
bipolar wind are of order $10^{55}$ erg, in line with the energy
requirement of the MSX observations.  We infer that the Galactic Center
is driving large-scale winds into the halo every $\sim 10-15$~Myr or
so.  
\end{abstract}

\section{Introduction}

Nuclear starburst-driven winds are now observed in dozens of external 
galaxies across the electromagnetic spectrum (Strickland 2001; Martin
1999; Heckman 2002; Veilleux 2002). For nearby galaxies, detailed
studies reveal that the superwind energetics are in the range 10$^{53}$
to 10$^{56}$ erg.  Galaxy-scale winds are suspected to be important in
both galaxy evolution and for the state of the intergalactic medium. In
the Galaxy, early enrichment of the halo and the outer disk through
powerful nuclear winds may account for the observed abundances of the
thick disk and globular clusters (Freeman \& Bland-Hawthorn 2002).
Moreover, nuclear winds can regulate the growth of the central bulge
(Carlberg 1999) and moderate the evolution of dwarf satellites
(Irwin\etal\ 1987).

At a distance of only 8.0$\pm$0.5 kpc (Reid 1993), the Galactic Center
shows a remarkable range of energetic activity at infrared, radio,
x-ray and $\gamma$-ray wavelengths (Morris \& Serabyn 1996 -- hereafter
MS; Yusef-Zadeh\etal\ 2000; Cheng\etal\ 1997). While this
activity has proved difficult to disentangle, there is now solid
evidence on scales of arcminutes to tens of degrees for powerful mass
ejections from the Galactic Center.

Potential sources of energy are young star clusters or the $2-3\times
10^6$ M$_\odot$ central black hole (Oort 1977; Frogel 1988; Genzel\etal\
1996; Ghez\etal\ 1998).  Individual star clusters have ages ranging
from 5~Myr (Krabbe\etal\ 1995) to 20~Myr (Eckart\etal\ 1999;
Figer\etal\ 2000; Gezari\etal\ 2002).  While the star formation history
is undoubtedly complicated, there is now abundant evidence that the
Galactic Center has experienced starburst episodes at different times
in the past (e.g.  Tamblyn \& Rieke 1993; Sjouwerman\etal\ 1998;
Simpson\etal\ 1999).  Detailed modelling of the PAH and fine structure
features (Lutz 1998) suggests a starburst event some 7~Myr ago,
supported by a census of individual stars (Genzel\etal\ 1994; 
Krabbe\etal\ 1995; Najarro\etal\ 1997).

An acute shortcoming of galaxy wind studies to date has been deriving
reliable energetics from multi-wavelength observations. We will argue 
that the proximity of the Galactic Center makes this one of the most 
important arenas for studying the wind process.  But current estimates
of the central energetics span a huge range.  Sofue \& Handa (1984)
presented evidence of a radio lobe on scales of $\sim 200$ pc (now
referred to as the Galactic Center Lobe) with implied thermal
energetics of roughly $3\times 10^{51}$ erg. Chevalier (1992) argued
for a higher energy content ($\sim 2\times 10^{52}$ erg) arising from
stellar winds driven by hot young stars over a 30 Myr duration.
Several authors have deduced that an explosive nuclear event
($4-8\times 10^{53}$ erg) must have taken place from the high
temperatures implied by the ASCA detection of 6.7~keV K$\alpha$ line
emission from He-like Fe~XXV (Koyama\etal\ 1996; Yamauchi\etal\ 1990).
On larger scales, Sofue (1977; 1984; 1994; 2000) has long argued that
the North Polar Spur is the byproduct of a central explosion which
took place 15 Myr ago. In his model, the scale of the blast is 10$-$20
kpc, with explosive energies in the range 10$^{55}$ to 3$\times
10^{56}$ erg.

The source of fuel for the observed activity appears to be the central
molecular zone (CMZ), a molecular `ring' with radius 180~pc and total
mass M$_{\rm cmz} \sim 8\times 10^6$ M$_{\odot}$.  Recent estimates of
mass inflow to the Galactic Center are of order 1 M$_\odot$ per year
(MS).   This is sufficient to trigger starbursts and nuclear activity
associated with powerful Seyfert galaxies ($\sim$10$^{43}$ erg s$^{-1}$).
The kinematics suggest the gas is kept on $x_2$ orbits by a strong
central bar, which guarantees cloud-cloud collisions on timescales of
$2\pi/\omega_c \approx 10^8$ yrs.  Of particular interest here, a
central explosion of order $10^{55-56}$ erg would provide a central gas
mass M$_{\rm cmz}$  with sufficient radial impulse to produce the
observed ring (Sanders 1989; Saito 1990).  Sophisticated hydrodynamical
simulations (e.g.  Suchkov\etal\ 1996; Strickland \& Stevens 2000) show
that such rings are indeed formed by central explosions and that disk
material is entrained by the expanding wind.

So what are the true energetics of the Galactic Center wind, and is it
a periodic or continuous flow? Is the wind driven by the central black
hole or by bursts of star formation, or both?  Here, we present the
first clear evidence that the nuclear wind which gave rise to the
Galactic Center Lobe (GCL) is entraining large amounts of molecular gas
and dust in the flow. The newly derived energetics, which are much
larger than earlier estimates, are of the same order needed to explain
the North Polar Spur (NPS) in terms of a central explosion, which
strongly suggests that the Galactic Center undergoes periodic bursts of
nuclear activity, as we show.

In \S~2, the new MSX observations are presented and compared to
existing far-infrared and radio data. In \S~3, we find that the
elevated dust temperature in the GCL requires a local heat source. In
\S~4, we derive the basic properties of the nuclear wind. In \S~5, we
revisit Sofue's earlier wind model for the NPS in light of new
observations.  In \S~6, we argue that the Galactic Center arena is of
extreme importance in our quest to understand large-scale winds.

\section{Observations} 

\begin{figure}
\vskip 4.8in
\includegraphics{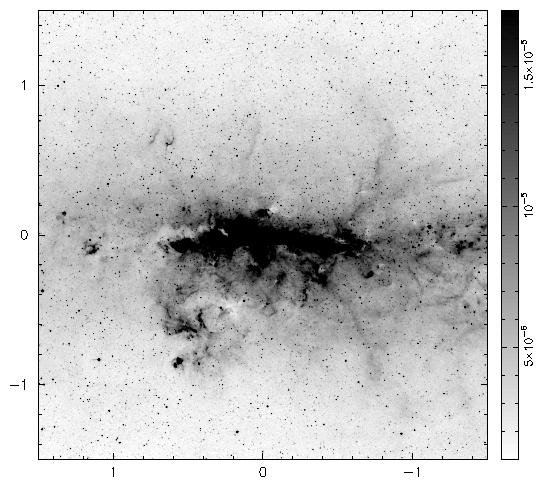}
\caption{\label{rawdata}
MSX image at 8.3$\mu$m of the Galactic Center presented in galactic
coordinates. The field of view is $\pm$1.5\deg\ in latitude and longitude
with the Galactic Center (defined by Sgr A$^*$) at the center of the field.
The bipolar dust shell is easily visible particularly on the western side
both above and below the Galactic Plane. The radiance units in the intensity
scale are W m$^{-2}$ sr$^{-1}$.
}
\end{figure}

In Fig.~\ref{rawdata}, we present a 3\deg\ $\times$ 3\deg\ region of
the Galactic Center reconstructed from scans with the MSX satellite in
`Band A' at 8.28$\mu$m. The Galactic Center is at the center of the
field.  The same data are presented in Fig.~\ref{unsharp} after
unsharp masking to accentuate the dust above the Galactic Plane (see below).
The MSX image is made up of 6\arcsec\ pixels generated by convolving
position-tagged, artefact-corrected Level 2A data. The adopted kernel
was gaussian with a 3\arcsec\ standard deviation using a Cartesian
projection and Galactic coordinates for the world coordinate system.
This results in a resolution of about 20\arcsec.  Full details of the
Galactic Plane survey, data processing and the creation of the images
are given by Price\etal\ (2001).  The image is calibrated absolutely in
radiance units of W m$^{-2}$ sr$^{-1}$, and a mosaic of 3$\times$3
individual MSX images was created to span a 3\deg\ $\times$
3\deg\ area.  

In Fig.~\ref{rawdata}, the central disk is characterized by point
sources (stars), small extended regions (compact and ultracompact
\HII\ regions), but the overwhelming impression is of a lattice of
overlapping filaments and bubbles along the Galactic Plane. Many of
these features have counterparts in radio continuum (Cohen \& Green
2001; Mezger \& Pauls 1979).  Compared to the IRAS 12$\mu$m data, the
MSX 8.3$\mu$m map is four times more sensitive, with 35 times higher
areal resolution. Note from the star counts that there is moderate
extinction in the plane even at 8$\mu$m.

Our particular interest lies in the spectacular mid-infrared `spurs' which
rise 1\deg\ above and below the plane at $\ell \approx -0.7$\deg. There
are similar features, rather less well defined, at $\ell \approx
+0.2$\deg.  The prominent western spur first came to light in the
original 20~cm scans of Kerr \& Sinclair (1966).  In Figs.~\ref{radio}
and \ref{unsharp},
we have superimposed the 3~cm map of Sofue (1985) which has a beam size
of 4.3\arcmin\ (HPBW).  There is a remarkably tight correlation of
off-planar structure in both the radio and mid-infrared maps, in
particular the eastern and western spurs at 3~cm coincide precisely
with the same features at 8$\mu$m. The spurs, which are also evident at
6~cm and 11~cm, appear to be largely thermal in origin, although there
is some evidence for synchrotron emission from the eastern spur
(Inoue\etal\ 1984).

\section{The Galactic Center Lobe}

\subsection{Dust temperature and emissivity}

The western spur is prominent in the IRAS data base (MS)
in the 12, 25, 60, 100$\mu$m bands.  In a 2\arcmin\ beam, 
Gautier\etal\ (1984) obtained a dust temperature of $T_d = 27\pm1$~K at
$\ell = 359.6$\deg, $b = 0.84$\deg. This temperature is broadly
consistent with the dust temperature maps of Schlegel, Finkbeiner \& 
Davis (1998).  We now determine whether the dust temperature,
or the observed constant emissivity of the spurs vertical to the plane, 
are consistent with heating by the stellar radiation field.

The total flux at a frequency $\nu$ reaching a dust cloud located
at a distance $r$ is obtained from integrating the specific
intensity $I_\nu$ over an optically thin disk with uniform emissivity,
\begin{equation}
F_\nu = \int I_\nu({\bf n})({\bf n}.{\bf N})\ d\Omega
\end{equation}
where ${\bf n}$ and ${\bf N}$ are the directions of the line of sight
and the outward normal to the surface of the disk, respectively.
For the general case, if we consider the flux seen at an arbitrary point
along the polar axis of the disk,
\begin{equation}
F = {L_{\rm bol}\over{\pi R^2}} (1-{1\over{\sqrt{1+(R/r)^2}}})
\end{equation}
where the bolometric luminosity of the Galaxy $L_{\rm bol}$ is
integrated over frequency out to a radius $R$. In the far field ($r \gg
R$), this equation reduces to the familiar form $F = L_{\rm bol}/2\pi
r^2$. In the near field ($r < R/2$), the flux along the polar axis is
constant. In principle, this could explain why the mid-infrared
emissivity along the spurs is roughly constant.

The equilibrium temperature a dust grain will assume at a distance $r$
from the source is given by 
\begin{eqnarray}
T_d &=& \left({{\varepsilon_a}\over{\varepsilon_r}}\ {{L_{\rm bol}}\over{2\pi\sigma r^2}}\right)^{0.25} \ \ \ \ \ \ \ \ \ \ \ \ \  \ \ \ \ \ \ \ \ \ \ \ \ \ \ \ \ \ {\rm far\ field} \\
&=& \left({{\varepsilon_a}\over{\varepsilon_r}}\ {{L_{\rm bol}}\over{\pi\sigma R^2}}(1-{1\over{\sqrt{1+(R/r)^2}}})\right)^{0.25}\ \ {\rm near\ field}
\end{eqnarray}
where ($\varepsilon_a,\varepsilon_r$) are the absorption and re-radiation
efficiencies, and $\sigma$ is the Stefan-Boltzmann constant.

If we assume that $\varepsilon_a \approx \varepsilon_r$, then for R $=$
12 kpc and $L_{\rm bol} = 3^{+2}_{-1} \times 10^{10}$ L$_\odot$, we
derive $T_d = 5$~K at a vertical distance of 100~pc. For an exponential
disk with scale length R$_e$ $=$ 3.5~kpc, the dust temperature rises to
$T_d = 9$~K.  The infrared spurs appear to require a much larger
heating rate than what the disk can provide.  Possible mechanisms are 
shock heating from the wind, or turbulent heating from the entrainment 
process. These processes are discussed below.

\subsection{Shell properties}

In Figs.~\ref{rawdata} -- \ref{unsharp},  we show for the first time that
dust is clearly associated with the {\it entire} GCL structure. This is
a particularly important result since it shows that the Galactic Center
wind is entraining dust in the flow. There is widespread evidence of
entrainment in external galaxy winds. This includes reversals of the
magnetic field in wind-blown shells, where the gas also shows evidence
for rotation and expansion (Cecil\etal\ 2001; 2002). Furthermore, hot
wind gas detected in [O${\scriptstyle\rm VI}$] is found to be moving
more rapidly than the cool gas (Heckman\etal\ 2001).

The measured column density of dust in the shell\footnote{The IRAS 
observations are largely sensitive to cool dust; future mid-infrared 
studies may reveal hotter dust in the shell but we expect that this 
contributes only a negligible fraction of the total dust content.} 
implies a hydrogen column density of N$_{\rm sh} \approx 2\times
10^{22}$ H$_2$ cm$^{-2}$ which compares with N$_{\rm sh} \approx
3\times 10^{22}$ H$_2$ cm$^{-2}$ determined from CO observations by
Sofue (1995) when averaged over both spurs.  These high column
densities could lead to even lower dust grain temperatures in the shell
than derived in \S~3.1 due to preferential absorption of high energy
radiation. This is not guaranteed since the gas and dust could have a
complex (fractal) distribution.  We expect that the shell comprises a
complex matrix of molecular, neutral and ionized gas.  There are
numerous nebular sources within the ISM for which the dust temperature
measured independently from, say, the warm ionized medium and warm
neutral medium are essentially identical (Lagache\etal\ 2000;
Heiles\etal\ 2000), suggesting some form of local thermal equilibrium
between the different phases.

We now estimate the mass of entrained material in the wind.
We model the GCL as a `telescope dome'-shaped shell with thickness
$\delta$. If the cylindrical part has a height $h$ and radius $r$,
the total volume of the bipolar shell is $V_s = 4\pi r \delta (r+h)$.
The volume enclosed by the dome in both hemispheres is $V_d = 2\pi r^2
(0.67 r + h)$.  In order to reproduce the 8.3$\mu$m observations, our
preferred model is ($\delta$,r,h) = (5,60,110) pc which gives $V_{\rm
sh}$ $\approx$ $2\times 10^{61}$ cm$^3$ and $V_d$ $\approx$ $9\times
10^{61}$ cm$^3$.  The expected limb-brightening contrast is ${\cal C}
\approx 2(\delta/{\cal B})^{0.5} (r/{\cal B})^{0.5}$ where ${\cal B}$
is the beam size of the IRAS measurement.  Thus, we find ${\cal C}
\approx 7$ which explains why the central bipolar cavity looks
remarkably free of filamentation compared to outside this region.

In our model, for the derived value of N$_{\rm sh}$ (\S 3.1), the
volume-averaged shell density is $n_{\rm sh} \approx 150$ H$_2$
cm$^{-3}$.  Therefore, the total mass of the shells in both hemispheres
is roughly M$_{\rm sh} \sim 5\times 10^6$~M$_\odot$. For a shell expansion
velocity of $v_{\rm sh}$ \kms\ (Sofue 1985), then the inferred {\it
minimum} energetics in the bipolar wind is $E_{kin} \sim 1\times
10^{54}$ $(v_{\rm sh}/100)^2$ erg.  New observations by Sofue (1996)
appear to indicate that the shell density could be a factor of a few lower
($n_{\rm sh} \sim 100$ cm$^{-3}$), so a more conservative estimate of
the required energetics is $\sim 3\times 10^{53}$ $(v_{\rm sh}/100)^2$
erg.

The vertical velocity component of the shell structure is presently
uncertain: we consider $v_{\rm sh} = 150$\kms\ to be a plausible estimate.
Large radial and vertical motions are apparent in about a third of the CMZ
gas (Bally\etal\ 1988). The high-velocity component has a velocity range
of 130--200\kms\ but this is driven in part by the bar. We conclude that
the minimum energy requirement for the dense shell is about 10$^{54}$ erg
for a covering fraction of unity (\ie\ assuming a closed shell surface).

\section{Starburst-driven winds.}

\subsection{Introduction}

Starburst and/or AGN-driven winds are observed in several dozen
external galaxies. Detailed numerical simulations shed light on the
basic mechanism (Strickland \& Stevens 2000; Tomisaka \& Ikeuchi 1988;
Suchkov\etal\ 1994). In nuclear starbursts, vast amounts of energy are
injected by supernovae and stellar winds into a small central volume
($<$ 100 pc). The `injection zone' reaches temperatures of about
10$^8$~K before the enormous pressure begins to drive a flow outwards.
The nuclear wind is initially subsonic but becomes supersonic at the
edge of the injection zone (Chevalier \& Clegg 1985). An adiabatic wind
accelerates as it expands into the clumpy ISM (Williams \& Dyson 2002)
producing a conventional forward shock, contact discontinuity and
reverse shock.  The hot nuclear wind is surrounded by a weaker wind
which percolates laterally through the clumpy medium. The hot
wind drags material in a shear flow along the contact discontinuity;
the weaker wind entrains advected cool material along with the flow.
The effectiveness of winds in entraining material is unclear, but
models suggest moderate (Strickland \& Stevens 2000) to large
(Suchkov\etal\ 1996) amounts of mass loading can take place.

It was initially thought that the bipolar x-ray emission seen in
starburst galaxies could be explained by cooling in an adiabatic wind
(Fabbiano 1988; Bregman\etal\ 1995). McCarthy\etal\ (1987) attempted
to confirm Chevalier \& Clegg's wind model of M82
by assuming the optical filaments were in thermal equilibrium with
the adiabatic wind.  However, the ram pressure of the  wind pressure
greatly exceeds the thermal pressure once the wind breaks free of the
injection zone.  Strickland\etal\ (1997) find that the observed x-ray emissivity
in M82 declines too slowly to be explained by an adiabatic wind.  Both the optical
and x-ray emission are more likely to arise from ram pressure confinement
which produces a much flatter fall off with radius, and can even lead
to increasing emissivity with radius (Veilleux\etal\ 1994).

Due to the proximity of the Galactic Center, we make the simplifying
assumption that all phases of the ISM are uniformly filled at the
resolution of the data (cf. Williams \& Dyson 2002). The explicit
dependence of derived quantities on filling factors can be found in
Cecil\etal\ (2001; 2002).

\subsection{Wind properties}
We now derive the properties of a galactic wind required to drive the
shell material.  ASCA has provided hard evidence for hot plasma towards
the Galactic Center (\S 1), and this is seen on similar scales
to the GCL.  The detection of very intense 6.7 keV K$\alpha$ line from
He-like Fe (Koyama\etal\ 1989; Yamauchi\etal\ 1990) leads to temperature
estimates of 10$^7$ to 10$^8$K. The temperature range implies {\it
initial} wind speeds ($\approx \sqrt{2\gamma k T_o / m_H}$) in the range
$v_o \approx 1700-3000$\kms\ as expected from the numerical models.
The Fe K$\alpha$ from the Galactic Center correlates well with the
dense gas, which is expected since the hot x-rays are easily reflected
by the cold neutral medium.  Once the x-ray spectra are corrected for
the scattered component, the implied temperatures appear to be at the
lower end of this range.

The diameter of the injection zone is expected to be of order 100~pc in
size. The dusty shell requires an energy input of at least 10$^{54}$
erg.  The clumpy appearance of the thermal radio emission and dust
emission suggest that the covering fraction may be $\kappa \lta 0.1$,
such that the required energetics are $E_{\rm min} \gta 10^{55}$ erg.
This is in line with Sofue's (2000) minimum requirement to explain the
North Polar Spur.  The density of the hot medium over the dome volume
is $E_{\rm min}/(3 k T_o V_d)$ which is of order $n_o \sim 1$ cm$^{-3}$
giving a mass of roughly 10$^5$~M$_\odot$ in the wind, where we have
adopted the high temperature limit in line with simulations.  For the
inferred optical depth to the Galactic Center (A$_v$ $\approx$ 30;
Becklin\etal\ 1978), x-ray absorption will only be significant for
lower temperature gas.  The expected x-ray emission measure is $2
\int^r_o n_e^2\ d\ell\ \sim\ 100$ cm$^{-6}$ pc.

In order to arrive at a very rough dynamical timescale, let us treat
the GCL as a pressurized bubble with radial dimension $r$, initially
located at a radius $r_*$ within the stellar bulge, which is lifted by
buoyancy.  (This requires that the wind is only now breaking out of 
the injection zone for the first time.) Assume that the medium
outside of the shell has to fall around the lobe at a speed $v_{\rm FF}
\sim \sqrt{G M_*(r_*) r/r_*^2}$ and that the lobe must move $2 r$. The
buoyancy timescale is therefore $\tau_b \sim 2 r_* \sqrt{ r/G
M_*(r_*)}$ yr. Our picture is of a light bubble that is
temporarily overpressured relative to its surroundings and thus rises
and expands as the ambient density falls. The enclosed mass in the
Galactic Center is $M_* \sim 3\times 10^6$~$r_*$~M$_\odot$, where $r_*$
is specified in parsecs (MS), which leads to $\tau_b \sim 1-2$~Myr.
McCray (1987) gives an approximate form for the cooling function
$\Lambda(T_e)$, from which we derive the cooling time in the hot wind,
$\tau_o \sim 2 k T_o / (\Lambda(T_o) n_o) \sim 10^8$ yr.  This greatly
exceeds the buoyancy timescale:  the nuclear wind cannot cool
radiatively and must therefore expand adiabatically.

For the derived wind model, the required pressure $P_o/k \sim 6\times
10^7$ K cm$^{-3}$ is an order of magnitude higher than what is possible
with stellar winds from the central regions (Chevalier 1992).
Remarkably, the derived properties ($n_o, T_o$) are characteristic of
{\it individual} supernova remnants (Holt \& McCray 1982).  Since the
volume of the GCL ($3\times 10^6$ pc$^3$) is at least three orders of
magnitude larger than a typical supernova remnant, this testifies to a
very large energy injection into a smaller central volume at some time
in the past.

\subsection{Wind-driven shocks}
\begin{figure}
\vskip 6.5in
\includegraphics{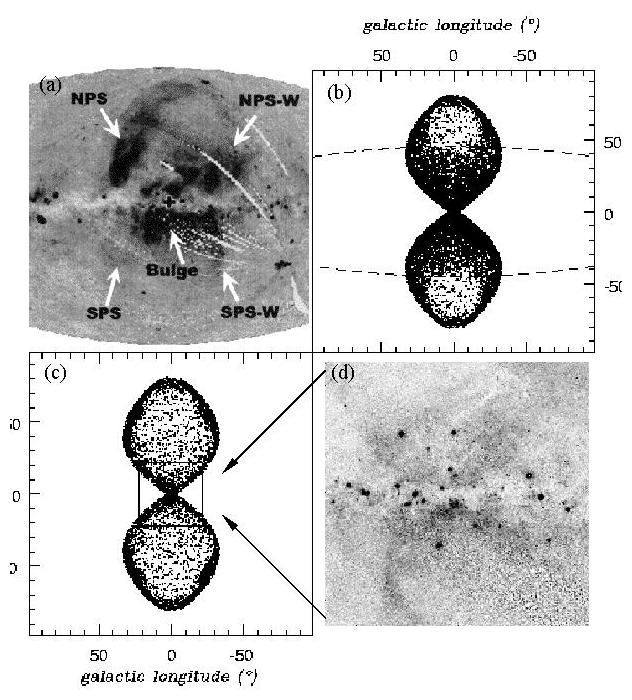}
\caption{\label{4panel}
(a) ROSAT 0.75~keV image in galactic coordinates centered on the Galactic 
Center. The field shown extends over $\pm$90\deg\ in latitude and longitude. 
The labels are defined in Sofue (2000). (b) The total projected emission
presented in Fig.~3(b) over the same angular scale as (a). Inside of the 
dashed ellipse, the x-ray emission
is complicated by attenuation, and by bulge and disk sources. (c) The
projected emission presented in Fig.~3(c). The box corresponds to the 
field of view of the ROSAT 1.5~keV image in (d).  Bipolar x-ray emission is 
clearly seen $\pm$20\deg\ above and below the Galactic Plane.
}
\end{figure}

We observe soft x-ray emission along the polar axis of the Galaxy out
to at least 2~kpc. The ROSAT 1.5~keV image presented by Wang (2002)
clearly shows the hourglass geometry characteristic of a bipolar flow
(see Fig.~\ref{4panel}(d)).
This emission is very unlikely to arise from thermal cooling in the wind.  
A more likely explanation is that the fast wind (Koo \& McKee 1992) is 
driving shocks into the halo gas via ram pressure. The energy of
the ROSAT band would implicate temperatures of $2\times 10^7$~K, or
shock velocities $v_s \sim 1200$\kms. But the initial shock velocity
at the base of the flow may be somewhat lower at $v_s \sim 600$\kms.
When a shock is driven into dense gas, the reflected shock re-heats
the post-shocked gas to even higher temperatures. This phenomenon has 
now been resolved in the Cygnus Loop by the Chandra x-ray satellite
(Levenson\etal\ 2002; Hester\etal\ 1994).

If we assume that all of the thermal energy in the injection zone is
converted to kinetic energy, the initial wind speed is in excess of
$v_w \sim 2000$\kms.  The manner in which the wind parameters vary with
radius depends in part on the wind geometry (Strickland\etal\ 1997).
For an adiabatic wind, the mass density and temperature decline as $\rho
\propto (r_o/r)^2$ and $T \propto (r_o/r)^{2(\gamma-1)}$ where $\gamma
= 5/3$. Thus, at the 2 kpc radius of the ROSAT emission, the wind 
temperature has dropped to 10$^6$~K and the density to $n_w \sim
10^{-3}$ cm$^{-3}$ where we have adopted $r_o = 100$~pc.

Therefore, at 2~kpc, if the visible shocked material is ram-pressure
confined, the pre-shock density is roughly $n_1 \sim n_w (v_w/v_s)^2
\sim 10^{-2}$ cm$^{-3}$. The post-shock material will be heated to a
temperature of $T_2 = 1.4 \times 10^5 (v_s/100)^2 \approx 5\times
10^6$K for singly shocked material (Hollenbach \& McKee 1979), or even
higher temperatures for material heated by a reflected shock
(Hester\etal\ 1994).  The post-shock temperature $T_2$ is within the
range favoured by numerical models (e.g.  Suchkov\etal\ 1994),
providing the gas has roughly solar abundances.  We have assumed a
fixed wind speed although in practice the wind could accelerate along
the flow, depending on the wind geometry.  In either case, the
post-shock densities will be of order $n_2 \sim 0.1$ cm$^{-3}$, much
less than that inferred from the dusty shell.  This is in line with the
density of material responsible for the x-ray emission in the
hydrodynamical models (Suchkov\etal\ 1994; Strickland\etal\ 1997).

The post-shock zone ($n_2,T_2$) has two important consequences. First,
it is the expected source of the soft x-ray emission observed along the
minor axis (\S 5).  The limb-brightened emission measures will
be 10$^{-2}$ or higher, in agreement with Sofue's hypershell model.
This assumes that the attenuation by the Galactic ISM is low above $b =
10$\deg, which appears to be so (Sofue 2000).

Secondly, for a sufficiently soft radiation field ($\lta 0.7$~keV),
we show below that the expected UV/x-ray emission is partly 
soaked up by the dense shell downstream of the shock.  This creates an
ionized interface between the hot wind and the cool gas in the shell.
A partially ionized region must exist in order to produce the thermal
radio emission although the required electron densities are about
$5-10$ cm$^{-3}$ (Sofue \& Handa 1984; Sofue 1985).  Once the radio
background is subtracted, the lobe has a flat (`thermal') spectral
index, $f_\nu \propto \nu^{-0.1}$, where $f_\nu$ is the radio source
flux. This behaviour is not normally associated with a shell-type
supernova remnant, and may reflect self-ionization by the shock-induced
UV field (Sutherland \& Dopita 1993), providing Fermi acceleration is
unimportant and the wind is non-relativistic.  The derived mass of the
thermal plasma is M$_{\rm th} \sim 4\times 10^5$ M$_\odot$. A rough
estimate of the ionization fraction is therefore M$_{\rm th}/$M$_{\rm sh}$
$\sim$ 10\%.

The UV/x-ray emissivity of the post-shock region may account for much
of the local ionization and heating, particularly if the shock velocity
does not exceed 600\kms\ or thereabouts.  The expected emissivity per
unit solid angle is $\rho_2 v_2^3 / 4\pi \sim 10^{-3}$ erg cm$^{-2}$
s$^{-1}$ sr$^{-1}$ where $\rho_2 = m_H n_2$ and $m_H$ is the mass of
the H atom.  If the shell is completely filled, most of the photons
will interact with the bipolar shell; half of the photons will
contribute to line radiation, and the remainder will produce energetic
secondary electrons.  If $\sim$30~eV is given up in each ionization by
secondary electrons, we derive an H$\alpha$ emission measure of
$\sim$100 cm$^{-6}$ pc, and a local electron density of about $n_e \sim
10$ cm$^{-3}$. This suggests the bipolar shell is partially ionized
with $n_e/n_{\rm sh} \sim 10\%$ in agreement with the rough value
derived above.  This agreement may be fortuitous: the covering fraction
{\it seen by the local radiation field} may be 10\% or less which
will lower the derived H$\alpha$ emission measure and ionization
fraction.

The derived x-ray flux is too low to irradiate the dust grains to the
observed temperature (\S 3.1; Voit 1991). An alternative heat source
are the Ly$\alpha$ photons produced in the post-shock region which are
continuously absorbed and re-emitted by neutral atoms until they are
ultimately absorbed by dust. If the Ly$\alpha$ heating is balanced by
the far-infrared emission, an equilibrium grain temperature can be
derived. Following Spitzer (1978), the derived grain temperatures fall
in the range 20--50~K depending on the details of the grain composition
(Draine \& Lee 1984). The contribution to the grain heating by electron
collisions within the shell gas is expected to be small.  Collisional
heating requires low gas densities ($\ll 1$ cm$^{-3}$) and much higher
kinetic temperatures ($\gg 10^5$K), but such harsh environs can easily
destroy grains (Dwek \& Arendt 1992). We refrain from presenting a more
detailed analysis since the post-shock environs are expected to be very
complicated for dust grains (McKee\etal\ 1987).

\section{The North Polar Spur revisited}

\subsection{Introduction}

The North Polar Spur extends 80\deg\ in galactic latitude perpendicular
to the plane, and is the largest projected structure seen at x-ray and
radio wavelengths (Berkhuijsen 1972; Haslam\etal\ 1974;
Snowden\etal\ 1997; Sofue 2000). In a long running series of papers,
Sofue (1977; 1984; 1994; 2000) has suggested that the NPS can be
explained in terms of an extremely powerful central explosion in the
last 15~Myr with a blast radius reaching to 10 kpc or more.  

More local origins have been proposed (\qv\ Egger \& Aschenbach 1995).
However, we agree with Sofue that a Galactic Center origin is more
likely for a number of reasons: (i) the radio continuum in the NPS is
thermal rather than the synchrotron emission observed locally in
supernova remnants; (ii) the loop appears well aligned with the
Galactic Center; (iii) there is now well established evidence on scales
of arcminutes to tens of degrees for powerful mass ejections from the
Galactic Center; (iv) the required energetics may be rather lower
($\sim 10^{55}$ erg) than the model favoured by Sofue -- such energies
are relatively common in well studied, nearby galaxies (Veilleux
2002).

In light of the new results presented here, we now revisit Sofue's
(2000) hypershell model for the North Polar Spur.  We return to the
discussion of the wind parameters derived in \S 4.3.  At 10 kpc, the
wind density has declined to about 10$^{-4}$ cm$^{-3}$ and the wind
temperature to about 10$^5$~K. The expected emission measures in the
post-shocked gas are now a hundred times lower than what we inferred at
a radius of 2~kpc which falls below the sensitivity limit of x-ray
satellites and so would not be seen.  How are we to understand this? A
possible explanation is that the post-shock densities are in fact
higher than we derive, due to material dragged along or accelerated by
the wind, and being shocked by the wind. This is a highly complex
process that has not been adequately modelled to date.

\subsection{A new perspective}

Another interpretation, which we now discuss in detail, considers how
the hypershell model\footnote{We use Sofue's term `hypershell' to refer
to our configuration of a central cone evolving to a cylinder at large
vertical distances above the plane.} would appear {\it in the near
field}. This step improves on Sofue's model in two ways. The NPS is
seen to extend to galactic latitudes of 80\deg, and therefore any
large-scale model {\it must} consider the angular projection.  First, our
model provides a simple explanation of why the NPS appears closed at
the top, an effect that is not apparent in Sofue's model. Second, the
near field projection provides a very long pathlength through the
hypershell on the near side, such that even a very low emission measure
can be boosted by factors of a hundred or more. As a result, the
required nuclear energetics of the wind can be much lower (see below).

We point out that Sofue (2000) did carry out a near-field projection
of a conical wind flow but the $\Omega$-shaped geometry arises from a
truncation along the vertical axis at a constant wind density. Sofue's
particular cone model does not lead to limb brightening at the apex of
the NPS, and in fact leads to caustics at $b = 90$\deg\ which are
inconsistent with the x-ray/radio observations.

Consider the hypershell geometry in Fig.~\ref{model}(a).
Narrow cone angle or cylindrical outflows are evident in
several starburst galaxies (Bland \& Tully 1988; G\"otz\etal\ 1990;
McKeith\etal\ 1995; Shopbell \&
Bland-Hawthorn 1998) and arise in hydrodynamical models (Tomisaka \&
Ikeuchi 1988; Strickland \& Stevens 2000).\footnote{A spectacular
example of the hypershell morphology is provided by the Hubble Heritage
image of the bipolar planetary nebula M2-9 at {\tt
heritage.stsci.edu/2001/05/mz3/m2-9.jpg}} The hydrodynamical models
reproduce this combination of an open central cone developing into a
more restricted opening angle further out.

The projected morphology in the model over the inner 30\deg\ is {\it
precisely} what is observed in the 1.5 keV ROSAT survey data (see
Fig.~\ref{4panel}(d) taken from Wang 2002). The x-ray emission extends
16\deg\ in elevation above and below the Galactic Center.  The model
presents the hypershells in both hemispheres. Sofue (2000) finds only
weak evidence for the southern hypershell on the largest scales, but we
note here that the 1.5 keV ROSAT survey clearly reveals the southern
biconal outflow.

Now consider how the hypershell geometry would appear {\it in the near
field}.  Recall that the NPS extends to 80\deg\ in galactic latitude.
Our viewpoint in the Galactic Plane is indicated by the symbol $\oplus$
in Fig.~\ref{model}(a).  The near-field projection is shown in
Fig.~\ref{model}(b).  The open hypershell now {\it appears} to close since
our line of sight at high latitude intersects a large column through
the biconal shell. The observed `mirage' is quite robust to a range of
parameters, although it does require that the central cone turns
inwards inside of the Solar Circle, and that the geometry becomes
cylindrical or even tapers (\ie\ the wine glass effect).  

There are two important reasons why the hypershell geometry must be
confined to the Solar Cylinder (\ie\ the cylinder which extends 
perpendicular to the Galactic Plane and intersects it
at the Solar Circle). First, if the conic
geometry continues to extend beyond the Solar Circle rather than turn
inwards, the projected emission produces caustics near $b=90$\deg.
Secondly, if the NPS is at the distance of the Galactic Center, the
projected angular width of the hypershell on the sky is fully
consistent with the turn up of the cone in the orthogonal plane defined
by $b = 0$\deg\ (compare Figs.~\ref{model}(a) and (b)).  This of course
assumes that the hypershell has cylindrical symmetry about its long
axis.  Given that the 1.5 keV data reveal a biconal geometry
extending above and below the plane by 2~kpc or more, it is difficult
to imagine another configuration that makes more sense given that it
{\it must} include the angular projection.  Our model is only
illustrative. We do not optimize the parameters since physical insight
requires a detailed hydrodynamical simulation matched to the observed
Galaxy parameters.

Figs.~\ref{model}(c) and \ref{model}(d) show the contributions of the
x-ray emission from the near and far side of the Galactic Center.  A
key prediction of our model is that much of the emission at low
latitude ($b < 50$\deg) within the NPS must arise from the far side of
the hypershell. The compression induced by the projection will partly
compensate for the inverse square dilution and absorption by the ISM.
The illustrative model shown here does not include attenuation by the
ISM (see Sofue 2000).  The limb effect arises from the nearside, and
the filling in arises largely from the farside of the hypershell. If we
add attenuation by intervening gas, it is possible to suppress the
degree to which projected shell appears filled in.

We now return to the original question in \S~5.1. The near-field effect
assists the interpretation of the x-rays in two ways. The source of the
emission is now $5-10$ kpc rather than $10-20$ kpc in Sofue's model;
this increases the wind pressure by a factor of $(15/7.5)^{2\gamma}$.
The increased post-shock density and the longer column through the
hypershell raises the emission measure by a factor of 100 to roughly
10$^{-2}$~cm$^{-6}$~pc or higher. This is Sofue's requirement in order
to match the NPS at 0.75~keV. In our model, a central explosion of $\sim
10^{55}$ erg, which was needed to explain the GCL, is probably
sufficient to account for the NPS, rather than the upper limit of
$3\times 10^{56}$ erg preferred by Sofue (2000).

We note with interest that the projected emission measure of the bipolar
shell at large radius would not be directly detectable in x-rays in
external galaxies. It is plausible that most disk galaxies with active
star formation have low surface-brightness bipolar winds. A physical
manifestation of this may be the extraordinary ionization conditions
that have been observed in the gaseous haloes of edge-on spiral
galaxies (\eg\ Collins \& Rand 2001). 

Our proposed model invites a range of tests.  In our model, the NPS is
very sharp edged which is expected regardless of how the hypershell
emission decays with vertical height above the plane. The angular limb
brightening caused by the near-field effect is more rapid than limb
brightening in a distance source.  Thus, there should be little hot gas
outside of the loop which defines the NPS.  Moreover, there
should be a lot of x-ray emission at galactic latitudes below $b =
50$\deg\ from the far side of the hypershell.  Furthermore, in line
with the hydrodynamical simulations, the gas on the inner edge of the 
NPS will be characteristically hotter and more redshifted (\ie\ moving
at higher velocities) than the gas towards the outer edge of the NPS.

Sembach\etal\ (1997) have shown the enormous potential of
absorption-line UV spectroscopy towards halo stars and active galaxies
used as background light sources.  The initial work with FUSE and the
Hubble Space Telescope reveals complex ionization and kinematics
towards 3C~273 whose sightline falls close to the NPS.  Halo clump
giants and blue horizontal branch stars are crucial for testing the
wind model. Their projected density at $V=17$ is about 1 per square
degree and they can be seen to 20~kpc.  An exciting prospect is the
NASA small explorer mission SPIDR\footnote{See
http://www.bu.edu/spidr/noflash/overview.html} due to launch in 2005.
This will map the distribution of CIV and OVI ionization over a quarter
of the sky (including NPS) at arcmin resolution.  Future observations
with SPIDR may reveal complex ionization and kinematics along and
across the NPS due to the turbulent boundary layer expected at the
wind-halo gas interface. The halo probes will need to have distances of
10~kpc or more in order to see this effect.

In summary, we propose that the North Polar Spur arises from limb
brightening as we look along the outer walls of the bipolar shell from
a near-field perspective.\footnote{This is easily verified with a
fluted wine glass initially observed at arm's length which is viewed
progressively closer until the vessel is just above eye level.}
Furthermore, the faint hypershell would be very difficult to detect in
external galaxies: {\it energetic bipolar winds may be far more common
in normal galaxies than is currently observed}.

\section{Review}

\begin{figure}
\vskip 6.5in
\includegraphics{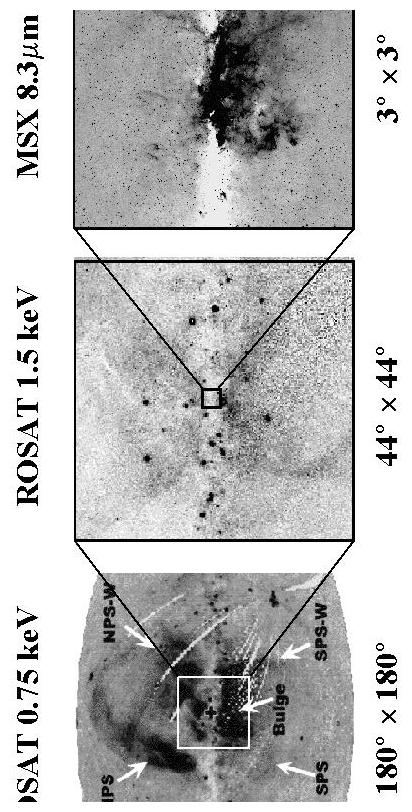}
\caption{\label{3panel}
A collage showing evidence for mass ejections on scales of degrees to
tens of degrees. The MSX image has been smoothed and unsharp masked to
emphasize the bipolar shells which are particularly prominent on the
west side.
}
\end{figure}

\smallskip\noindent{\bf Recurrent nuclear activity.}
We have presented evidence for mass ejections from the Galactic Center
on scales of a few degrees to tens of degrees. This is evident from
extended bipolar emission at mid-infrared, radio and x-ray wavelengths;
the three very different scales are shown together in Fig.~\ref{3panel}.
The inner shell visible in the 8.3$\mu$m data, and physically
associated with the Galactic Center Lobe, has a dynamical timescale of
about a million years.  This compares with roughly 15~Myr for the
hypershell model of the North Polar Spur. On all scales, the observed
structures show a remarkable similarity to the observed behavior in NGC
3079 (Cecil\etal\ 2001; 2002) where an inner radio-emitting shell is
observed at the centre of a much large conic shell. We note that NGC
3079 and the Galaxy share important similarities in their wind
parameters and their galaxy properties.  

\begin{figure}
\vskip 6.5in
\includegraphics{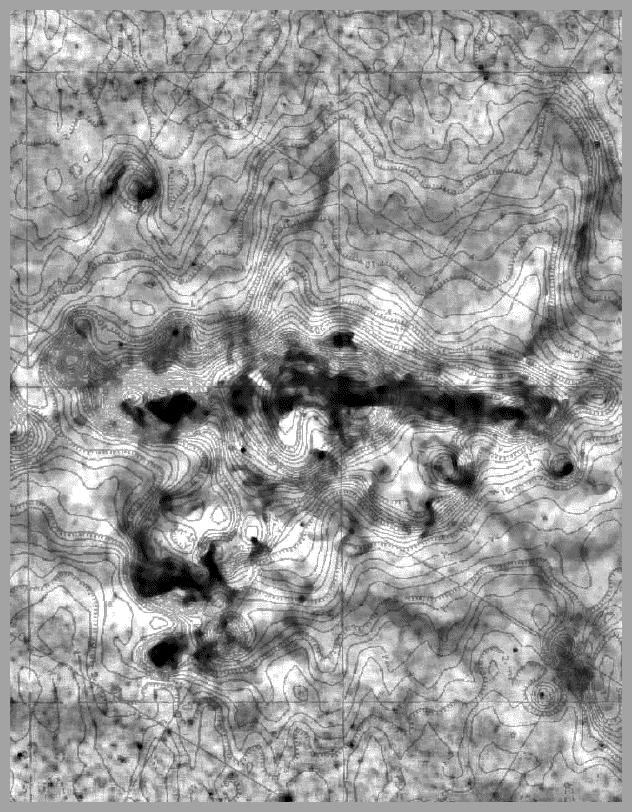}
\caption{\label{unsharp}
An unsharp masked version of Fig.~2 to accentuate the faint dust
emission at the northern tip of the GCL. The 3~cm radio contours
are superimposed to emphasize the correspondence between the ionized
gas and the faintest dust emission far from the plane. For the NE, NW
and SW spurs, the dust appears confined to the outside of the ionized. 
}
\end{figure}

Moreover, note that the bipolar geometry and the central star forming
region in Fig.~\ref{unsharp} appear tipped over by about 20\deg\ 
towards the west with respect to the Galactic Plane, a phenomenon that 
is seen in M82 (Shopbell \& Bland-Hawthorn 1998).  This
may arise from a tilt of the orbital plane of the central gas and/or
stellar bar with respect to the Galactic Plane, for which there is some
evidence (MS).

The different dynamical timescales of the GCL and NPS indicate that
these arise from different events. However, we find in \S~5.2 that both
phenomena require a central event with roughly equivalent energetics of
about 10$^{55}$ erg, and this activity must be periodic on timescales
of 10$-$15~Myr.  The stellar census provides clear evidence of several
starbursts which must have occurred in this time frame, and these are
the likely source of the energy input.  It is not guaranteed that each
starburst must produce an energetic wind.  If the supernovae largely
occur within dense gas, their thermalization effiencies can be
$\varepsilon \ll 0.1$ such that the explosive energy is radiated away
(Thornton\etal\ 1998).  The timescale of the GCL could be consistent
with evidence pointing to a major starburst in the last 7~Myr. For an
energy injection zone of 100 pc or so, the wind may not develop for a
few million years after the event. This has been investigated by
sophisticated numerical simulations (Suchkov\etal\ 1994; Strickland \&
Stevens 2000; Strickland\etal\ 2002).

\smallskip\noindent{\bf The resolution problem.}
We stress that studies of the large-scale wind in the Galaxy are of
fundamental importance in understanding this complex process. To
understand why, we return to the powerful nuclear wind (or superbubble)
in NGC 3079. High resolution x-ray and optical images
(Strickland 2001; Cecil\etal\ 2002) show that most of the mass of the
wind resides within the extended filaments, and that the wind material
is clumpy down to the smallest scales observed. The smallest cloud
scales in simulations must be very well sampled to treat entrainment
and acceleration properly (Klein, McKee \& Colella 1994), but the
necessary resolution is one or two orders of magnitude smaller than
what can be presently achieved (Strickland 2001).  In NGC 3079, we may
be seriously underestimating the mass loss and associated energetics in
the nuclear wind.

The Galactic Center does not suffer resolution problems to anything
like the same degree, and is therefore an extremely important source of
information on the ubiquitous phenomenon of nuclear winds. Although the
extinction to the Galactic Center from x-ray to mid-infrared
wavelengths is high, moderate levels of extinction are also inferred
towards the nuclear winds in external galaxies because these need to be
highly inclined in order to see the extended wind emission. 

The proximity of the Galactic Center suggests a number of important
experiments.  The observed $^{26}$Al line emission in $\gamma$-rays
(Mahoney\etal\ 1984; Genzel \& Townes 1987) indicates that supernova 
activity appears to be ongoing. Sub-millimeter observations will
determine whether these supernovae reside within dense or tenuous gas.
We have identified starburst activity as the likely source of the wind,
but it is well known that AGNs produce powerful winds (\eg\ Schiano
1986). A key experiment is to trace the $^{26}$Al $\gamma$-ray line
emission or the pressure profile in thermal radio emission to establish
if the injection zone is resolved or continues to rise towards the
central black hole. Radio continuum measurements are also needed in
order to measure the transverse magnetic field associated with the
wind.  Magnetohydrodynamical processes can radically alter the
development of the wind.  The post-shock regions are likely to be
complicated.  The FUSE, Chandra and XMM satellites are essential for
measuring gas temperatures where the wind interacts with the halo or
entrained ISM, either through direct emission or from absorption lines
towards background sources.  In the coming years, we anticipate a far
deeper understanding of galactic winds arising from detailed
multiwavelength studies of the Galactic Center.

\begin{figure}
\vskip 4.8in
\includegraphics{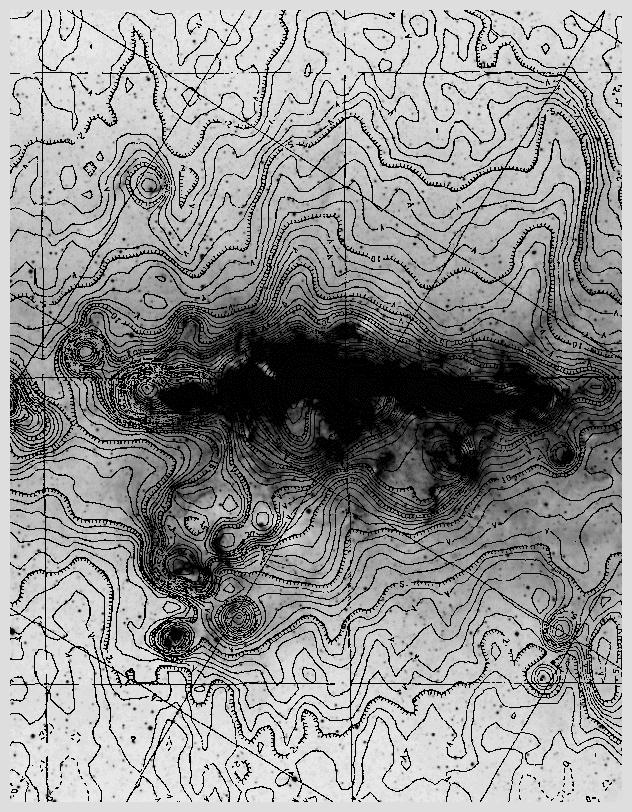}
\caption{\label{radio}
Fig.~1 with the 3~cm radio contours superimposed. Note the
close correspondence between the dust and ionized gas in the
western spurs. The correspondence for the NE, NW and SW spurs is 
particularly clear with increasing vertical height from the plane. 
The SE spur is much less clear in both the MSX and 3~cm data.
}
\end{figure}

\smallskip
We are indebted to David Strickland and Eli Dwek for invaluable
comments, and to Gerald Cecil and Andrew McGrath for help with
overlaying the radio map in Fig.~\ref{radio}.  This research made use
of data products from the Midcourse Space Experiment.  Processing of
the data was funded by the Ballistic Missile Defense Organization with
additional support from NASA Office of Space Science.  This research
has also made use of the NASA/ IPAC Infrared Science Archive, which is
operated by the Jet Propulsion Laboratory, California Institute of
Technology, under contract with the National Aeronautics and Space
Administration.  

\begin{figure}
\vskip 5.5in
\includegraphics{fig4.ps}
\caption{\label{model}
(a) An illustrative bipolar shell (hypershell) model for the Galaxy
seen in the far field.  Our viewpoint at the Solar Circle is shown by 
the symbol $\oplus$. The other figures (b)--(d) are the {\it same}
hypershell geometry seen {\it in the near field}, \ie\ after projecting
(a) as seen from $\oplus$ into proper galactic coordinates. (b) is
the predicted total emission, (c) is emission from the near side of the
hypershell, (d) is emission from the far side as defined by the plane
of the sky through the Galactic Center. 
}
\end{figure}

\end{document}